\documentclass{article}

\usepackage{graphicx}
\usepackage[round]{natbib}
\usepackage{epsfig}

\title{
Objective Probabilistic Forecasts of Future Climate Based on Jeffreys' Prior: the Case of Correlated Observables
}

\author{Stephen Jewson\footnote{\emph{Correspondence email}: \texttt{stephen.jewson@rms.com}},
Dan Rowlands, Myles Allen\\}

\begin{document}
\maketitle

\begin{abstract}
To include parameter uncertainty into probabilistic climate forecasts one must first specify a prior.
We advocate the use of objective priors, and, in particular, the Jeffreys' Prior.
In previous work we have derived expressions for the Jeffreys' Prior for the case in which the observations
are independent and normally distributed. These expressions make the calculation of the prior much
simpler than evaluation directly from the definition.
In this paper, we now relax the independence assumption and derive expressions for the Jeffreys' Prior for the case in which the
observations are distributed with a multivariate normal distribution with constant covariances. Again, these
expressions simplify the calculation of the prior: in this case they reduce it to the calculation
of the differences between the ensemble means of climate model ensembles based on different parameter settings.
These calculations are simple enough to be applied to even the most complex climate models.
\end{abstract}

\section{Introduction}

Predicting the future climate is very difficult:
this can be seen in the wide spread of predictions and projections that come from different climate prediction models~\citep{ipcc4}.
This wide spread is indicative of forecast uncertainty,
and for those who might use climate predictions, it is important that this uncertainty is quantified as well as possible.
Only then can decisions be made as to whether certain climate predictions should be used or ignored, and, if they are to be used,
how they should be used and how much weight they should be given.

One major driver of climate prediction uncertainty is parameter uncertainty: that is, that different parameter settings in climate
models lead to different predictions, and that no single set of parameters is correct.
In statistical climate predictions, parameter uncertainty can be estimated and incorporated rather
easily using standard statistical techniques.
For instance, \citet{j115} discusses a Bayesian method for putting parameter uncertainty into
predictions from flat-line and linear trend climate prediction models.
In numerical model climate predictions, however, parameter uncertainty is rather more difficult to estimate and incorporate
into predictions since the models
are more complex and the number of parameters is much larger. Some attempts to deal with numerical model parameter uncertainty
using classical statistics are described in \citet{allen09}, and some that use subjective Bayesian statistics are described in \citet{tomassini07}.
We, however, are particularly interested in the idea of using a third statistical paradigm known as
\emph{objective} Bayesian statistics to capture the parameter uncertainty in numerical climate models.
Objective Bayesian statistics is Bayesian statistics in which those priors which cannot be determined from
previous independent studies are determined using a rule, rather than subjectively using intuition. We have described how
this approach can be applied to climate models in \citet{jp1}. We proposed using the Jeffreys' prior~\citep{jeffreys}
which is the most standard of the various rules available, and
we then showed that the implementation of such an approach can be simplified by using a parametric
form for the predictive distribution from the climate model.
This simplifies the calculation of the prior, because the two steps of differentiating the probabilities from the climate
model and taking the expectation over all possible realisations are replaced by a single step of differentiating the parameters
of the fitted distribution.

One of the shortcomings of the method described in~\citet{jp1} is that we assumed that the observables (or predicted variables) were
independent. This simplifies matters because it means that the predictive distribution factorises into a product of predictive
distributions for each variable. Since we assumed a multivariate normal predictive distribution, that factorised into a product
of independent normal distributions, and the evaluation of the prior became a matter of differentiating the means and the variances
of those normal distributions.

In this paper, we stick with the assumption that the observations are multivariate normally distributed. However, we now relax the assumption that the observations are independent, and replace it with the assumption that the covariance matrix between the observations is constant as a function of the climate model parameters. This allows us to derive another set of relatively simple expressions for the prior, that could readily be evaluated using a suitably designed set of integrations of a numerical model.

In section~\ref{s2} we give a brief overview of objective priors and the work in \citet{jp1}.
In section~\ref{s3} we then derive expressions for the new case we consider in this paper, for the constant variance multivariate normal.
In section~\ref{s4} we summarise.

\section{The Use of Objective Priors in Climate Modelling}\label{s2}

The starting point for Bayesian methods for making probabilistic forecasts that include parameter uncertainty is the following equation
from probability theory, sometimes known as the law of total probability:

\begin{equation}
p(y|x)=\int p(y|\theta) p(\theta|x) d\theta
\end{equation}

This equation says that the probability of future event $y$, given the past data $x$, is given by
an average over probabilistic predictions made with all possible parameter values,
where the prediction for $y$ from a model based on the parameter value $\theta$ is written as $p(y|\theta)$.
The average is a weighted average, where the weights on each prediction are given by the likelihood $p(\theta|x)$.

Thus far, there is nothing Bayesian about this equation, since Bayes' Rule has not yet been used.
In Bayesian methods the term $p(\theta|x)$ is evaluated by factorising it using Bayes' Theorem:
\begin{equation}
    p(\theta|x) \propto p(x|\theta) p(\theta)
\end{equation}

The task of evaluating $p(\theta|x)$ now becomes a question of evaluating $p(x|\theta)$, which is the
probability of the past events $x$ given each parameter value $\theta$, and of evaluating the prior $p(\theta)$.

Evaluating $p(x|\theta)$ is, at least conceptually, straightforward since it involves comparing probabilistic
predictions from a climate model with past observations.
The distinction between subjective and objective Bayesian methods appears in how the term $p(\theta)$ is
evaluated. In subjective methods $p(\theta)$ is based on intuition, while in objective methods it is based on a rule.
The idea of using a rule is to make the resulting predictions less arbitrary: this is discussed at greater length in~\citet{jp1}.

The most widely discussed rule is the Jeffreys' Prior, given by:
\begin{equation}\label{eq1}
    p(\theta)=\sqrt{\mbox{det}\left[-\mbox{E}\left(\frac{\partial^2 \log p(x|\theta)}{\partial \theta_j \partial \theta_k}\right)\right]}\\
    \label{jp_def}
\end{equation}

If this equation is applied to all parameters, then there is a standard case in which the results are questionable.
This problem is widely discussed in textbooks on Bayesian statistics (for instance, see page 90 in~\citet{peterlee}).
The problem was actually resolved, however, by Jeffreys' himself (see page 1345 of~\citet{kasswasserman} for an explanation),
using the separate treatment of location parameters.

In~\citet{jp1} we discussed how Jeffreys' Prior might be evaluated for a climate model.
The steps in equation~\ref{eq1} in which the log of the probability are first differentiated and then integrated (to evaluate the expectation)
are somewhat daunting, and likely to be computationally intensive.
However, they can be simplified if we make the assumption that the output from the model
is independent and normally distributed.
The steps of differentiation and integration can then be performed analytically, and the expression above becomes:
\begin{equation}\label{eq2}
p(\theta)=\sqrt{\mbox{det}\left(\sum_{i=1}^{n}
 \frac{2}{\sigma_i^2}\frac{\partial \sigma_i}{\partial \theta_j}\frac{\partial \sigma_i}{\partial \theta_k}
+\frac{1}{\sigma_i^2}\frac{\partial \mu_i}   {\partial \theta_j}\frac{\partial \mu_i}{\partial \theta_k}
\right)}
\end{equation}
where $n$ is the number of observations used to validate the model, and $\mu_i$ and $\sigma_i^2$ are the means and variances
of initial condition ensembles.
To evaluate this new expression, the means and variances from initial condition ensembles need to be differentiated with respect
to the underlying parameters of the climate model $\theta$.
One can imagine doing this by running multiple initial condition ensembles (although there may be more efficient methods).
This is still not a trivial exercise, but is simpler than differentiating the probabilities from the model
and integrating over all possible realisations.
Also, with careful experimental design it should be possible to evaluate this expression by re-using the model integrations that are
needed to calculate the likelihood term $p(x|\theta)$.

In the case that the ensemble variance does not vary as a function of the parameters, expression~\ref{eq2} reduces to:
\begin{equation}\label{eq6}
p(\theta)=\sqrt{\mbox{det}\left(\sum_{i=1}^{n}
\frac{1}{\sigma_i^2}\frac{\partial \mu_i}   {\partial \theta_j}\frac{\partial \mu_i}{\partial \theta_k}
\right)}
\end{equation}

\section{Jeffreys' Prior for Correlated Observations with Constant Variance}\label{s3}

We now consider a slightly different approximation in order to allow for correlations between observations.
We still assume that the observations come from a multivariate normal distribution, as before.
However, instead of assuming that the observations are \emph{independent}, but that the ensemble variance can \emph{vary} as a function of the
parameters, we now make a complementary set of approximations in which we assume that the observations are \emph{correlated},
but that the covariance matrix is \emph{constant} as a function of the parameters.

In general, probability densities from the multivariate normal distribution are given by:
\begin{equation}
p(x|\theta)=\frac{1}{(2\pi)^\frac{n}{2}} \frac{1}{D^{\frac{1}{2}}} \mbox{exp}\left(-\frac{1}{2}(x-\mu)^T S(x-\mu)\right)
\end{equation}

where
\begin{itemize}
  \item $x$ is a vector for the observables, or predicted variables, produced by the model
  \item $\theta$ is a vector for the parameters in the model
  \item $\mu$=$\mu(\theta)$ is a vector of the mean response of the model for each observable ({i.e.} the ensemble mean
  of $x$ for an infinite-sized initial condition ensemble for fixed parameters $\theta$)
  \item $\Sigma=\Sigma(\theta)$ is the covariance matrix of the response of the model between observables
   ({i.e.} the ensemble covariance matrix of $x$ for an infinite-sized initial condition ensemble for fixed parameters $\theta$)
  \item $S=\Sigma^{-1}=S(\theta)$ is the inverse of the covariance matrix
  \item and $D=\mbox{det}(\Sigma)=D(\theta)$ is the determinant of the covariance matrix
\end{itemize}

This gives:
\begin{eqnarray}
\ln p(x|\theta)
&=&-\frac{n}{2}\ln 2\pi -\frac{1}{2} \ln D -\frac{1}{2}(x-\mu)^T S(x-\mu)\\
&=&-\frac{n}{2}\ln 2\pi -\frac{1}{2} \ln D -\frac{1}{2}(x^T S x-x^TS \mu-\mu^T S x+\mu^T S \mu)
\end{eqnarray}

Since $S$ is symmetric, $x^T S \mu=\mu^T S x$ (see appendix 1)
which means that the above expression for $\ln p(x|\theta)$ simplifies a little to:
\begin{eqnarray}
\ln p(x|\theta)
&=&-\frac{n}{2}\ln 2\pi -\frac{1}{2} \ln D -\frac{1}{2}(x^T S x-2x^TS \mu+\mu^T S \mu)
\end{eqnarray}

We now consider two cases.

\subsection{Single parameter, multiple correlated observations}

The first case we consider, as a warm-up, is where there is just a single parameter in the climate model.

If we consider $\theta$ to be this single (scalar) parameter, then:
\begin{eqnarray}
\frac{\partial \ln p(x|\theta)}{\partial \theta}
&=&-\frac{1}{2} \frac{\partial \ln D}{\partial \theta}
   -\frac{1}{2}x^T \frac{\partial S}{\partial \theta} x
   +x^T\frac{\partial (S \mu)}{\partial \theta}
   -\frac{1}{2}\frac{\partial (\mu^T S \mu)}{\partial \theta}
\end{eqnarray}

We now make the further approximation that the covariance matrix $\Sigma$ is constant, which simplifies the subsequent algebra considerably.
If $\Sigma$ is constant, then both $D$ and $S$ are also constant, and so:
\begin{eqnarray}
\frac{\partial \ln p(x|\theta)}{\partial \theta}
&=& x^T S\frac{\partial \mu}{\partial \theta}
   -\frac{1}{2}\frac{\partial (\mu^T S \mu)}{\partial \theta}\\\label{eq3}
&=& x^T S\frac{\partial \mu}{\partial \theta}
   -\frac{1}{2}\left(\mu^T S\frac{\partial \mu}{\partial \theta}
   +\frac{\partial \mu^T}{\partial \theta}S \mu\right)
\end{eqnarray}

Again, since $S$ is symmetric,  $\mu^T S\frac{\partial \mu}{\partial \theta}=\frac{\partial \mu^T}{\partial \theta}S \mu$,
which simplifies the above expression to
\begin{eqnarray}\label{eq4}
\frac{\partial \ln p(x|\theta)}{\partial \theta}
&=&x^T S\frac{\partial \mu}{\partial \theta}-\mu^T S\frac{\partial \mu}{\partial \theta}
\end{eqnarray}

Taking another derivative wrt $\theta$ gives:
\begin{eqnarray}
\frac{\partial^2 \ln p(x|\theta)}{\partial \theta^2}
&=&x^T S\frac{\partial^2 \mu}{\partial \theta^2}-\mu^T S\frac{\partial^2 \mu}{\partial \theta^2}-\frac{\partial \mu^T}{\partial \theta} S\frac{\partial \mu}{\partial \theta}
\end{eqnarray}

Taking expectations over $x$ (and noting that $E(x)=\mu$) gives:
\begin{eqnarray}
E\left(\frac{\partial^2 \ln p(x|\theta)}{\partial \theta^2}\right)
&=&E\left(x^T S\frac{\partial^2 \mu}{\partial \theta^2}\right)-E\left(\mu^T S\frac{\partial^2 \mu}{\partial \theta^2}\right)-E\left(\frac{\partial \mu^T}{\partial \theta} S\frac{\partial \mu}{\partial \theta}\right)\\
&=&E(x^T) S\frac{\partial^2 \mu}{\partial \theta^2}-\mu^T S\frac{\partial^2 \mu}{\partial \theta^2}-\frac{\partial \mu^T}{\partial \theta} S\frac{\partial \mu}{\partial \theta}\\
&=&\mu^T S\frac{\partial^2 \mu}{\partial \theta^2}-\mu^T S\frac{\partial^2 \mu}{\partial \theta^2}-\frac{\partial \mu^T}{\partial \theta} S\frac{\partial \mu}{\partial \theta}\\
&=&-\frac{\partial \mu^T}{\partial \theta} S\frac{\partial \mu}{\partial \theta}
\end{eqnarray}

That the second derivatives cancel and disappear from this expression is no surprise: we know from a standard result that the Jeffreys' Prior can only
contain first derivatives (see lemma 2, page 87, \citet{peterlee}).

The Jeffrey's Prior is then given by:
\begin{equation}
p(\theta)=\sqrt{\frac{\partial \mu^T}{\partial \theta} S\frac{\partial \mu}{\partial \theta}}
\end{equation}

If there are $n$ observations then:
\begin{itemize}
  \item $\mu$ is an $n$ x $1$ vector
  \item $\frac{\partial \mu}{\partial \theta}$ is an $n$ x $1$ vector
  \item $\mu^T$ is a $1$ x $n$ vector
  \item $\frac{\partial \mu^T}{\partial \theta}$ is a $1$ x $n$ vector
  \item $S$ is an $n$ x $n$ matrix
  \item $\frac{\partial \mu^T}{\partial \theta} S\frac{\partial \mu}{\partial \theta}$ is a scalar
  \item and $p(\theta)$ is a scalar
\end{itemize}

In the case in which the observations are independent, $\Sigma$ reduces to a diagonal matrix with the variances of each
of the $n$
observations $\sigma_1^2,\sigma_2^2,...,\sigma_n^2$ on the diagonal, and
$S$ is a diagonal matrix with the inverse variances on the diagonal.
The matrix multiplication in the above expression then reduces to a simple sum over the observations:
\begin{eqnarray}
p(\theta)
&=&\sqrt{\frac{\partial \mu^T}{\partial \theta} S\frac{\partial \mu}{\partial \theta}}\\
&\propto&\sqrt{\sum_{i=1}^n \frac{\partial \mu_i}{\partial \theta} \frac{1}{\sigma_i^2}\frac{\partial \mu_i}{\partial \theta}}\\
&\propto&\sqrt{\sum_{i=1}^n \frac{1}{\sigma_i^2} \left(\frac{\partial \mu_i}{\partial \theta}\right)^2}
\end{eqnarray}

which agrees with the uncorrelated observations case given by equation 20 in~\citet{jp1}.

\subsection{Multiple parameter, multiple observations}

We now consider the more general case with multiple parameters.
As a first step we consider two parameters.
Taking the derivative of $\frac{\partial \ln p(x|\theta)}{\partial \theta}$, from equation~\ref{eq4} above, wrt a second parameter $\phi$ gives:

\begin{eqnarray}
\frac{\partial^2 \ln p(x|\theta)}{\partial \theta \partial \phi}
&=&x^T S\frac{\partial^2 \mu}{\partial \theta \partial \phi}
   -\mu^T S\frac{\partial^2 \mu}{\partial \theta \partial \phi}
   -\frac{\partial \mu^T}{\partial \phi} S\frac{\partial \mu}{\partial \theta}
\end{eqnarray}

Taking expectations gives:
\begin{eqnarray}
E\left(\frac{\partial^2 \ln p(x|\theta)}{\partial \theta \partial \phi}\right)
&=&E\left(x^T S\frac{\partial^2 \mu}{\partial \theta \partial \phi}\right)
   -E\left(\mu^T S\frac{\partial^2 \mu}{\partial \theta \partial \phi}\right)
   -E\left(\frac{\partial \mu^T}{\partial \phi} S\frac{\partial \mu}{\partial \theta}\right)\\
&=&E(x^T) S\frac{\partial^2 \mu}{\partial \theta \partial \phi}
   -\mu^T S\frac{\partial^2 \mu}{\partial \theta \partial \phi}
   -\frac{\partial \mu^T}{\partial \phi} S\frac{\partial \mu}{\partial \theta}\\
&=&\mu^T S\frac{\partial^2 \mu}{\partial \theta \partial \phi}
   -\mu^T S\frac{\partial^2 \mu}{\partial \theta \partial \phi}
   -\frac{\partial \mu^T}{\partial \phi} S\frac{\partial \mu}{\partial \theta}\\
&=&\frac{\partial \mu^T}{\partial \phi} S\frac{\partial \mu}{\partial \theta}
\end{eqnarray}

Generalising from two to multiple parameters, and writing the vector of parameters as $\theta$,
the Jeffreys' Prior is then:
\begin{equation}
p(\theta)=\sqrt{\mbox{det} \left( \frac{\partial \mu^T}{\partial \theta_j} S\frac{\partial \mu}{\partial \theta_k}\right)}
\end{equation}

If there are $n$ observations and $m$ parameters then:
\begin{itemize}
  \item $\mu$ is an $n$ x $1$ vector
  \item $\frac{\partial \mu}{\partial \theta}$ is an $n$ x $m$ matrix
  \item $\mu^T$ is a $1$ x $n$ vector
  \item $\frac{\partial \mu^T}{\partial \theta}$ is an $m$ x $n$ matrix
  \item $S$ is an $n$ x $n$ matrix
  \item $\frac{\partial \mu^T}{\partial \theta} S\frac{\partial \mu}{\partial \theta}$ is an $m$ x $m$ matrix
  \item det$\left(\frac{\partial \mu^T}{\partial \theta} S\frac{\partial \mu}{\partial \theta}\right)$ is a scalar
  \item and $p(\theta)$ is a scalar
\end{itemize}

In the independent observations case this reduces to:
\begin{eqnarray}
p(\theta)
&=&\sqrt{\mbox{det} \left( \sum_{i=1}^{n} \frac{\partial \mu_i}{\partial \theta_j} \frac{1}{\sigma_i^2} \frac{\partial \mu_i}{\partial \theta_k}\right)}\\
&=&\sqrt{\mbox{det} \left( \sum_{i=1}^{n} \frac{1}{\sigma_i^2} \frac{\partial \mu_i}{\partial \theta_j} \frac{\partial \mu_i}{\partial \theta_k}\right)}
\end{eqnarray}

which agrees with the uncorrelated observations case given by equation 37 in~\citet{jp1}, and is also given above as equation~\ref{eq6}.

\section{Summary and Discussion}\label{s4}

Bayesian statistics offers the only effective framework for including parameter uncertainty into probabilistic forecasting.
Part of the Bayesian approach involves specifying a prior distribution for the parameters, and there are two options for how to
do this: take a subjective approach, where the prior is based on intuition, and take an objective approach, where the prior is
based on a rule.

Of the various rules that might be used in the objective approach, one in particular is the most widely discussed, and is the closest
to being an accepted standard: the Jeffreys' Prior. When we consider how to apply Jeffreys' Prior in climate
modelling, we find that the probabilities from initial condition ensemble runs of a climate model need to be differentiated with respect
to the parameters of the model, and integrated over all possible model states. This could be done, but is cumbersome. However, by making
parametric assumptions for the shape of the predictive distributions there is the potential for this to be simplified.
The most obvious parametric assumption is that all output from the model is distributed according to the multivariate normal.
We have now considered two special cases of this. In~\citet{jp1} we considered the case where the observations were considered
to be independent, but where both the means and the variances of initial condition ensembles could vary as a function of the parameters.
In this paper we have considered the case where the observations are not independent, but where the covariance matrix between observations
is constant. Neither of these two cases is a special case of the other, clearly.

We have derived expressions for the Jeffreys' Prior for the constant covariance case.
These expressions show that evaluating the Jeffreys' Prior reduces to evaluating first derivatives of the ensemble mean with respect
to the various parameters in the model, and performing simple a calculation using those first derivatives.
This calculation involves the correlation matrix between different observations, which can be estimated from the grand ensemble
of all models runs with all parameter values, since we are assuming that the correlation matrix is constant.

Overall we have presented a method for making objective probabilistic forecasts
which is sufficiently simple that it could be applied to real climate models, even when the observations are considered correlated.
Our next challenge is to attempt to tackle the general multivariate normal case for correlated observations with non-constant covariance matrix,
and to consider predictive distributions other than normal.

\bibliography{arxiv}

\begin{thebibliography}{8}
\providecommand{\natexlab}[1]{#1}
\providecommand{\url}[1]{\texttt{#1}}
\expandafter\ifx\csname urlstyle\endcsname\relax
  \providecommand{\doi}[1]{doi: #1}\else
  \providecommand{\doi}{doi: \begingroup \urlstyle{rm}\Url}\fi

\bibitem[Allen et~al.(2009)Allen, Frame, Huntingford, Jones, Lowe, Meinshausen,
  and Meinshausen]{allen09}
M~Allen, D~Frame, C~Huntingford, C~Jones, J~Lowe, M~Meinshausen, and
  N~Meinshausen.
\newblock {Warming Caused by Cumulative Carbon Emissions Towards the Trillionth
  Tonne}.
\newblock \emph{Nature}, 458, 2009.

\bibitem[IPCC(2007)]{ipcc4}
IPCC.
\newblock Climate change 2007 - {T}he {P}hysical {S}cience {B}asis.
\newblock Technical report, IPCC Working group, 2007.

\bibitem[Jeffreys(1946)]{jeffreys}
H~Jeffreys.
\newblock {An Invariant Form for the Prior Probability in Estimation Problems}.
\newblock \emph{Proceedings of the Royal Society of London Series A},
  186:\penalty0 453--461, 1946.

\bibitem[Jewson(2008)]{j115}
S~Jewson.
\newblock {Weather Derivative Pricing and the Modelling of Trends: Objective
  Bayesian Versions of the Flat-line, Linear Trend and Damped Linear Trend
  Models}.
\newblock \emph{http://ssrn.com/abstract=1212523}, 2008.

\bibitem[Jewson et~al.(2009)Jewson, Rowlands, and Allen]{jp1}
S~Jewson, Dan Rowlands, and Myles Allen.
\newblock {A New Method for Making Objective Probabilistic Climate Forecasts
  from Numerical Climate Models Based on {J}effreys' {P}rior}.
\newblock \emph{arXiv:physics/0908.4207}, 2009.

\bibitem[Kass and Wasserman(1996)]{kasswasserman}
R~E Kass and L~Wasserman.
\newblock {The Selection of Prior Distributions by Formal Rules}.
\newblock \emph{Journal of the American Statistical Association}, 1996.

\bibitem[Lee(1997)]{peterlee}
Peter Lee.
\newblock \emph{Bayesian Statistics}.
\newblock Arnold, 1997.

\bibitem[Tomassini et~al.(2007)Tomassini, Reichert, Knutti, Stocker, and
  Borsuk]{tomassini07}
L~Tomassini, P~Reichert, R~Knutti, T~Stocker, and M~Borsuk.
\newblock {Robust Bayesian Uncertainty Analysis of Climate System Properties
  using Markov Chain Monte Carlo Methods}.
\newblock \emph{Journal of Climate}, 20:\penalty0 1239--1254, 2007.

\end{thebibliography}


\end{document}